\documentclass[aps,prb,showpacs,superscriptaddress,twocolumn,10pt,amssymb,amsfonts]{revtex4-1}

\usepackage{amsmath}
\usepackage{bm}
\usepackage{graphics}
\usepackage{dsfont}
\usepackage{subfigure}		
\usepackage{sidecap}
\usepackage{epsfig}
\usepackage[all]{xy}
\usepackage{hyphenat}
\usepackage{hyperref}

\newcommand{\bra}[1]{ {\langle{#1}|} }
\newcommand{\ket}[1]{ {|{#1}\rangle} }

\newcommand{\MxTwo}[2]{ {\begin{bmatrix} #1 \\ #2 \end{bmatrix}} }
\newcommand{\SmallMxTwo}[2]{ {\left[\!\begin{smallmatrix} #1 \\ #2 \end{smallmatrix}\!\right]} }

\newcommand{\Tr}[1]{ \operatorname{Tr}\left[{#1}\right] }

\newcommand{\vk}{ {\mathbf{k}} }
\newcommand{\vG}{ {\mathbf{G}} }
\newcommand{\va}{ {\mathbf{a}} }
\renewcommand{\vr}{ {\mathbf{r}} }
\newcommand{\vR}{ {\mathbf{R}} }
\newcommand{\ZZ}{ {\mathbb{Z}} }

\newcommand{\TR}{ \Theta }			
\newcommand{\Ta}{ T_{1/2} }		
\newcommand{\tildeTa}{ \tilde{T}_{1/2} }	

\begin{document}
\title{Antiferromagnetic topological insulators}
\author{Roger S.~K.~Mong}
	\affiliation{Department of Physics, University of California, Berkeley, CA 94720, USA}
\author{Andrew M.~Essin}
	\affiliation{Department of Physics, University of California, Berkeley, CA 94720, USA}
\author{Joel E.~Moore}
	\affiliation{Department of Physics, University of California, Berkeley, CA 94720, USA}
	\affiliation{Materials Sciences Division, Lawrence Berkeley National Laboratory, Berkeley, CA 94720}
\date{\today}

\newcommand{\essin}[1]{ {} }
\newcommand{\roger}[1]{ {} }
\newcommand{\joel}[1]{ {} }

\begin{abstract}

We consider antiferromagnets breaking both time-reversal ($\TR$) and a primitive lattice translational symmetry ($\Ta$) of a crystal but preserving the combination $S = \TR \Ta$.
The $S$ symmetry leads to a $\ZZ_2$ topological classification of insulators, separating the ordinary insulator phase from the ``antiferromagnetic topological insulator'' (AFTI) phase.
This state is similar to the ``strong'' topological insulator with time-reversal symmetry, and shares with it such properties as a quantized magnetoelectric effect.
However, for certain surfaces the surface states are intrinsically gapped with a half-quantum Hall effect [$\sigma_{xy} = e^2 / (2 h)$], which may aid experimental confirmation of $\theta = \pi$ quantized magnetoelectric coupling.
Step edges on such a surface support gapless, chiral quantum wires.  In closing we discuss GdBiPt as a possible example of this topological class.

\end{abstract}

\pacs{03.65.Vf, 75.50.Ee, 73.43.-f, 73.20.At, 85.75.-d}

\maketitle

\section{Introduction}

Several insulating materials are now known experimentally to have metallic surface states as a result of topological properties of the bulk electron wave functions.~\cite{HsiehTIDirac,HasanBi2Se3,ChenBi2Te3}  These ``topological insulators''~\cite{MooreTIBirth,HasanKaneReview} resulting from strong spin-orbit coupling were originally understood theoretically by classifying single\hyp{}electron states of materials with time\hyp{}reversal invariance,~\cite{FuKaneMeleTI3D,MooreBalents06,RoyQSH3D} building on previous work in the two\hyp{}dimensional (2D) case.~\cite{KaneMeleZ2}  The novel metallic surface states of the three\hyp{}dimensional (3D) topological insulators can be observed directly via angle-resolved photoemission spectroscopy (ARPES) and, in the simplest case, can be viewed as a reduced version of graphene with a single surface Dirac point, rather than two in the case of graphene, and a single spin state at each momentum rather than two.

Our goal in this paper is to explain how three\hyp{}dimensional antiferromagnetic insulators with \textit{broken} time-reversal symmetry can nevertheless have nontrivial features similar to that of the topological insulators.  Along some planar surfaces they have gapless surface modes, while along others the surface is gapped and there is a nonzero magnetoelectric coupling from the intrinsic material; an experimental signature in the latter case is the existence of one\hyp{}dimensional (1D) metallic states along step edges on the surface.  We concentrate here on the conditions for an antiferromagnetic insulator to be in the topologically nontrivial class and on the measurable consequences at its surfaces.

The time-reversal invariant topological insulators are described by $\ZZ_2$ topological invariants (\textit{i.e.}, there are only two possible values, ``odd'' and ``even'') that differ from the integer-valued topological invariants that underlie the integer quantum Hall effect (IQHE) in two\hyp{}dimensional time\hyp{}reversal\hyp{}breaking systems.  A simple picture of the state we discuss is obtained by starting from a nonmagnetic topological insulator on a bipartite Bravais lattice, then adding antiferromagnetic order that doubles the unit cell.  One of the three\hyp{}dimensional topological invariants survives in this process.  Note that this differs in several ways from the two\hyp{}dimensional model introduced by Haldane on the honeycomb lattice,~\cite{HaldaneQHE88} which is classified by the standard IQHE integer-valued topological invariant (TKNN integer~\cite{TKNN} or ``Chern number'') and where the time-reversal breaking does not change the structural unit cell, which is on the hexagonal Bravais lattice.  Another case previously considered is a system that breaks time\hyp{}reversal $\TR$ and spatial inversion $\Pi$ but preserves the combination $\TR \Pi$ (note that the Haldane model does not preserve this combination); here there are $\ZZ_2$ invariants in $d=1,2$ for spinless systems and no topological invariants for spin 1/2 systems.~\cite{RyuLudwigDimHeir}

The basic idea in this paper is to classify crystals with broken time-reversal $\TR$ but with an unbroken symmetry of the form $S = \TR \Ta$, where $\Ta$ is a lattice translation symmetry of the ``primitive'' (structural) lattice that is broken by the antiferromagnetic order.  Because the topological invariant involves explicitly the lattice operation $\Ta$, it is sensitive to how this lattice operation is modified by a surface, as mentioned above, and even the gapless surface state is not expected to be stable to disorder (in contrast to the conventional topological insulator).  A macroscopic description is useful in order to understand the conditions for the topological antiferromagnet to be stable.  The three\hyp{}dimensional topological insulator can be characterized by the existence of a quantized magnetoelectric coupling in the electromagnetic Lagrangian~\cite{QiHughesZhangTFT,WilczekAxion,EssinMPAxion} ($c=1$):
\begin{align}
	\Delta \mathcal{L}_\textrm{EM} = \frac{\theta e^2}{2\pi h} \mathbf{E} \cdot \mathbf{B},
		\quad \theta=\pi.
\end{align}
The coupling $\theta$ is only defined modulo $2 \pi$, and ordinary insulators with time-reversal invariance have $\theta = 0$.  The presence of either time-reversal symmetry or inversion symmetry is sufficient to guarantee that the other orbital magnetoelectric terms are absent.~\cite{EssinOMP,MalashevichOMP}  The product $S$ is also enough to guarantee that the space-averaged $\theta$ is quantized to zero or $\pi$, since $\theta$ is odd under $S$.

The bulk value $\theta=\pi$ allows either metallic surfaces or gapped surfaces, but in the gapped case there must be a half-integer quantum Hall effect.  In the conventional topological insulators, the surfaces are intrinsically metallic and observation of the magnetoelectric coupling seems to require adding a time\hyp{}reversal\hyp{}breaking perturbation.  In the topological antiferromagnets, some surfaces have a gapped state just from the material's own time\hyp{}reversal\hyp{}breaking, which suggests that experimental confirmation that $\theta=\pi$, which has not yet occurred, may be easier in these materials, using the same techniques previously used to extract $\theta$ in Cr$_2$O$_3$.~\cite{ObukhovSchmidExptTheta}  Surface disorder would complicate that approach but would enable observation of special features at step edges as discussed below.

In the following section, we define the topological antiferromagnet in terms of band structure and verify the connection to the macroscopic description in terms of magnetoelectric response.  Then the surface properties are discussed, which will likely be important for experimental detection.  In closing we discuss the requirements for experiment and comment on the possibility that the antiferromagnetic Heusler alloy GdBiPt may realize this phase;~\cite{CanfieldAFGdBiPt} the possibility that such Heusler alloys may include several topological insulators has recently been a topic of interest.~\cite{HasanHeusler,FelserHeusler}

\section{$\ZZ_2$ topological invariant}
\label{SectionZ2Inv}

In this section, we construct the $\ZZ_2$ invariant which
distinguishes between the trivial insulator and
``antiferromagnetic topological insulator'' (AFTI) phases.
We consider a antiferromagnet breaking both the primitive lattice
symmetry $\Ta$ and time-reversal symmetry $\TR$, but preserving the
combination $S = \TR \Ta$.
The unit cell is effectively doubled as a result and $\Ta^2$ is the
new lattice translation (which accounts for the notation).
In the following, lattice vectors are
elements of this doubled lattice except where otherwise specified.

A free particle Hamiltonian takes the form
$\mathcal{H} = \sum_{\vk\in \textrm{BZ}} \Psi_\vk^\dag H(k_1,k_2,k_3) \Psi_\vk$
in reciprocal space, where $\Psi^\dag$ and $\Psi$ are fermion
creation and annihilation operators; $k_1,k_2,k_3 \in [0,2\pi)$
are momentum coordinates defined by $k_i = \vk\cdot\va_i$; and
$\va_i$ are the lattice translation vectors. The eigenvectors
$u_\vk$ of the Hamiltonian $H(\vk)$ are related to the
wave functions by Bloch's theorem $\psi_\vk = e^{i\vk\cdot\vr}u_\vk$.
Consequently, the Hamiltonian is not periodic in $\vk$, but rather
satisfies $H(\vk+\vG) = e^{-i\vG\cdot\vr}H(\vk)e^{i\vG\cdot\vr}$,
where $\vG$ is a reciprocal lattice vector and $\vr$ is the position
operator in this context.
Finally, we single out $\va_3$ such that $\Ta^2$ gives a translation by $-\va_3$.

For spin-$1/2$ fermions, the time-reversal operator may be written as
$\TR = -i\sigma^y\mathcal{K}$ in a suitable basis,
where $\mathcal{K}$ is the complex conjugation operator.
In addition, $\TR$ (and $S$) also flips the sign of the momentum: $\vk \rightarrow -\vk$.
The translation operator $\Ta(\vk)$ will move
the lattice by half a unit cell, so that its representation in
reciprocal space satisfies $\Ta^2(\vk) = e^{ik_3}$.
Explicitly,
\begin{align}
\Ta(\vk) = e^{\frac{i}{2}k_3} \MxTwo{0 & \mathds{1}}{\mathds{1} & 0},
\end{align}
where $\mathds{1}$ is the identity operator on half the unit cell.
Note that the operators $\TR$ and $\Ta$ commute so that $\TR \Ta(\vk) = \Ta(-\vk) \TR$.

The combination $S_\vk = \TR \Ta(\vk)$ is antiunitary like $\TR$
itself, but with an important difference: while $\TR^2 = -1$ for
spin-1/2 particles, $S^2 = S_{-\vk}S_\vk = -e^{ik_3}$.
The Hamiltonian is invariant under the combination of time-reversal
and translation:
\begin{align}
	S_\vk H(\vk) S_\vk^{-1} = H(-\vk)
\end{align}
At the Brillouin zone (BZ) plane $k_3=0$ the Hamiltonian satisfies
$S H(k_1,k_2,0) S^{-1} = H(-k_1,-k_2,0)$ with  $(S|_{k_3=0})^2=-1$.
These properties lead to a $\ZZ_2$ topological classification of this
two\hyp{}dimensional system, by analogy to the $\ZZ_2$ invariant in
the quantum spin Hall (QSH) effect~\cite{KaneMeleZ2} (the same invariant can be rederived in the Hamiltonian picture used here).~\cite{MooreBalents06}
At the plane $k_3 = \pi$, by contrast, $S^2 = +1$ and there are no topological invariants associated with this plane.~\cite{SFRLClassification}

The $\ZZ_2$ invariant may be computed from the Berry connection and 
curvature~\cite{FuKaneTRPZ2,MooreBalents06} on the $k_3=0$ plane, or in the
presence of spatial inversion by looking at the four time-reversal momenta at
$k_1,k_2 \in \{0,\pi\}$.~\cite{FuKaneTIInversion}

Even though the topological invariant is calculated from a
two\hyp{}dimensional slice in the Brillouin zone for a particular choice of unit cell, it reflects the topology
of the three\hyp{}dimensional band structure. For example, $S$ symmetry gives no
invariants in 1D or 2D. In the Appendix, we show that the 3D $\ZZ_2$ invariant is independent of unit cell choice.
In the remainder of this section, we will give a more detailed picture of this topological phase.

\subsection{Relation to the time-reversal invariant topological insulator}
\label{SectionStrongTopIns}

If we imagine the system described by a time-reversal breaking order
parameter $M$ (\textit{e.g.}, a staggered magnetization), what happens when we
restore time-reversal symmetry by letting $M$ go to zero while
maintaining the insulating phase (band gap)?

To understand what happens, it is useful to recall briefly
the classification of three\hyp{}dimensional time\hyp{}reversal band
insulators.  In the Brillouin zone, there are six planes which
satisfy time-reversal $\TR H(\vk) \TR^{-1} = H(-\vk)$, and each has a
corresponding $\ZZ_2$ invariant:
$\alpha_0,\alpha_\pi,\beta_0,\beta_\pi,\gamma_0,\gamma_\pi$ classify
the planes $k_1=0,\pi$, $k_2=0,\pi$, and $k_3=0,\pi$ respectively.
Here, we use the convention 0 (even) and 1 (odd) to denote the elements of $\ZZ_2$.
The six values must satisfy the constraint $s \equiv
\alpha_0 + \alpha_\pi = \beta_0 + \beta_\pi = \gamma_0 + \gamma_\pi$,
all modulo 2, so
only four combinations of these quantities are independent:
$s, \alpha_0, \beta_0, \gamma_0$.  The value $s$ is the ``strong''
topological invariant, and the other three $\ZZ_2$ are known as the
``weak'' invariants; together they classify the 3D system.
A strong topological insulator (STI) is one in which $s$ is non-trivial,
that is, $s=1$.

Upon doubling the unit cell in the $\va_3$ direction, the
Brillouin zone halves by folding in the $k_3$ direction.  (In this
subsection only, $\va_3$ is the lattice vector of the structural
lattice, and $\va^d_3 = 2\va_3$ is the lattice vector of the
``doubled'' system which supports an antiferromagnetic coupling.)
We can write the Hamiltonian of the doubled system $H_\vk^d$ in terms of the undoubled Hamiltonian $H_\vk$:
\begin{align}
	H^d \left( k^d_3 \right)
		& = \mathcal{U} \MxTwo{ H(k^d_3/2) & 0 }{ 0 & H(k^d_3/2 + \pi) } \mathcal{U}^\dag
			\notag	\\
	\mathcal{U} & = \frac{1}{\sqrt{2}}
		\MxTwo{ \mathds{1} & e^{i\vG^d_3\cdot\vr} }{ \mathds{1} & -e^{i\vG^d_3\cdot\vr} }
\end{align}
Here 
$\vG^d_3$ is the reciprocal lattice vector dual to $\va^d_3$, 
$\vr$ is the position operator, and the dependence on $k_1$ and $k_2$ 
are omitted for brevity.  The unitary transformation
$\mathcal{U}$ ensures that the eigenvectors of $H^d$ satisfy
Bloch's theorem.

Under the doubling process, the $k_3=0$ and $k_3=\pi$ planes collapse
onto the $k^d_3=0$ plane.  The new invariant $\gamma^d_0$ is given
as a sum $\gamma_0 + \gamma_\pi = s$ since the unitary
transformation $\mathcal{U}$ does not affect any these topological
invariants.
On the other hand, the planes $k_3 = \pm\pi/2$ map
to the plane $k^d_3 = \pi$. Since $\pm\pi/2$ are time-reversal
conjugate and those planes (like all BZ planes, by assumption) have
vanishing Chern numbers, it can be seen that
$\gamma^d_\pi$ is always zero.

Adding an antiferromagnetic ($\TR$-breaking) parameter $M$ to a STI produces an AFTI.  Alternatively,
as we turn down the time-reversal breaking parameter $M$, the
antiferromagnet reverts to the doubled system. The $\ZZ_2$ invariant
describing our system is $\gamma^d_0 = s$ and we have a STI at $M=0$
(provided the bulk gap does not close).
This gives a natural way to construct a non-trivial
topological antiferromagnet - by taking a STI and introducing a
staggered magnetization which breaks time-reversal but preserves
$S$.

\subsection{Magnetoelectric effect and the Chern-Simons integral}
\label{MagnetoelectricCS}

The strong topological insulator exhibits a quantized magnetoelectric
effect, which can be taken as its definition.~\cite{WilczekAxion,QiHughesZhangTFT,EssinMPAxion}
To review briefly, the magnetoelectric response tensor
\begin{align}
	\alpha^i_j = \frac{\partial P^i}{\partial B^j} \Big\vert_{\mathbf{B}=0}
\end{align}
is odd under the action of time-reversal.  In a $\TR$-invariant medium,
this immediately restricts the off-diagonal elements of the tensor to
vanish.  However, the
ambiguity in defining the bulk polarization~\cite{KSmithVanderbilt,OrtizMartin94}
allows the diagonal elements to take a nonzero value.  In fundamental
units, the strong topological insulator has
\begin{align}
	\alpha^i_j = \frac{1}{2} \frac{e^2}{h} \delta^i_j
		= \frac{\theta}{2\pi} \frac{e^2}{h} \delta^i_j
\end{align}
with $\theta = \pi$.

The antiferromagnetic topological insulator suggested here does not
have time-reversal symmetry microscopically; the relevant symmetry
operation is $S$. 
This
distinction should not affect the macroscopic response of the system
to uniform fields (\textit{i.e.}, $\theta$), although there could be
short\hyp{}wavelength components of $\alpha^i_j$.

From the general theory of orbital magnetoelectric responses in band
insulators, the nonzero contribution to $\alpha^i_j$ in cases of discrete
symmetries such as time\hyp{}reversal comes from the Chern-Simons
integral,
\begin{align}
	\theta & = \frac{1}{4\pi} \int_\textrm{BZ}\! cs_3		\notag
	\\	cs_3 & = \Tr{A \wedge F + \tfrac{i}{3} A \wedge A \wedge A}
		\label{ChernSimons}
\end{align}
where $A^{\mu\nu} = \bra{u_\vk^\mu} id \ket{u_\vk^\nu}$ is the Berry connection
(a matrix\hyp{}valued 1-form), and $\mu, \nu$ label filled bands.~\footnote{Readers unfamiliar with differential forms, curvature, or Chern-Simons forms may consult M.~Nakahara, \textit{Geometry, Topology and Physics} (Taylor \& Francis, 1990) for reference.}
The curvature 2-form is $F = dA - iA\wedge A$.
Under a gauge transformation (a unitary transformation
between the bands), the Chern-Simons integral will change by an
integer multiple of $2\pi$,
hence only $\theta \bmod 2\pi$ is physical.

Under time-reversal $\ket{u_\vk} \rightarrow \TR\ket{u_\vk}$, the
quantities $\vk\rightarrow-\vk$ and $cs_3\rightarrow-cs_3$, and
$\theta$ changes sign.
The translation operator $\Ta = e^{\frac{i}{2}k_3}
\SmallMxTwo{0 & \mathds{1}}{\mathds{1} & 0}$ changes $cs_3$ by an exact form (total derivative) and does not affect $\theta$.  Together, $S$ symmetry implies that
$\theta = -\theta + 2\pi n$ for some integer $n$, which quantizes
$\theta$ to $0$ (topologically\hyp{}trivial phase) or $\pi$
(topological insulator phase).

The topological phase remains well-defined even when the single particle invariant is not, in the case with electron\hyp{}electron interactions.  The macroscopic $\theta$ angle remains quantized (at $0$ or $\pi$) as long as the bulk gap does not close, so the AFTI is stable to sufficiently weak interactions.

The presence of $S$ symmetry forces the Chern numbers on all BZ
planes to be zero.  In a three\hyp{}dimensional system, the three Chern
numbers
are the only obstruction to finding a set of continuous wave functions
in the Brillouin zone (respecting Bloch boundary conditions). This
guarantees the existence of a single-valued connection $A$ for
Eq.~\eqref{ChernSimons}. Such $A$ might not respect $S$ symmetry, but
this is no impediment to computing the Chern-Simons integral.

\section{Construction of effective Hamiltonian models}

In this section, we present two explicit examples of Hamiltonians in the antiferromagnetic topological insulator class. Henceforth, we refer these as ``model A'' and ``model B.''

\subsection{Construction from strong topological insulators}
\label{ModelSTI}

As noted in Sec.~\ref{SectionStrongTopIns}, we can create an
antiferromagnetic topological insulator by adding a staggered
time\hyp{}reversal breaking term to a strong topological insulator. Here
we present an explicit Hamiltonian constructed in such way.

We start with a four-band model on a cubic lattice by Hosur
\textit{et~al.},~\cite{HosurRyuChiralTISC} with four orbitals/spins per cubic site,
\begin{multline}
	H(k_x,k_y,k_z)
	\\	= v \tau^x \otimes (\sin k_x \sigma^x + \sin k_y \sigma^y +\sin k_z \sigma^z)
	\\	+ \big[ m + t(\cos k_x + \cos k_y + \cos k_z) \big] \tau^z,
\end{multline}
where $\bm{\sigma}$ and $\bm{\tau}$ are two sets of Pauli matrices.
This Hamiltonian is in the strong topological phase when
$|t| < |m| < 3|t|$ and $\lambda \neq 0$, with the
time-reversal operator represented by $-i\sigma^y \mathcal{K}$.

To double the Hamiltonian in the $z$ direction, first decompose
$H(k_x,k_y,k_z)$ into a hopping Hamiltonian as follows:
\begin{align}
	H(k_x,k_y,k_z) = B^0 + \sum_\mu \left( B_\mu e^{-ik_\mu} + B_\mu^\dag e^{ik_\mu} \right),
\end{align}
where $\mu = x,y,z$. The matrices $B_\mu$ describe hopping from
adjacent cells from the $-\mu$ direction, $B_\mu^\dag$ are hopping
from $+\mu$ direction, and $B^0$ describes the ``self\hyp{}interaction''
of a cell.
The new lattice vectors are:
\begin{align}
	\begin{bmatrix} \va_1 \\ \va_2 \\ \va_3 \end{bmatrix}
		= \begin{bmatrix} 1 & 0 & 1 \\ 0 & 1 & 1 \\ 0 & 0 & 2 \end{bmatrix}
			\begin{bmatrix} \va_x \\ \va_y \\ \va_z \end{bmatrix}
\end{align}
which defines a face\hyp{}centered\hyp{}cubic (FCC) lattice with the primitive unit cell whose volume
is double that of the original cubic cell.

\begin{widetext}
Doubling the unit cell gives the following eight-band Hamiltonian:
\newcommand{\khf}{ \frac{k_3}{2} }
\begin{multline}
	H^d(k_1,k_2,k_3) = \MxTwo{B^0 + M & B_z^\dag e^{i\khf}}{B_z e^{-i\khf} & B^0 - M}
		+ \MxTwo{0 & B_z}{0 & 0} e^{-i\khf} + \MxTwo{0 & 0}{B_z^\dag & 0} e^{i\khf}
	\\	+ \MxTwo{0 & B_x}{B_x & 0} e^{i(\khf-k_1)} + \MxTwo{0 & B_x^\dag}{B_x^\dag & 0} e^{i(k_1-\khf)}
		+ \MxTwo{0 & B_y}{B_y & 0} e^{i(\khf-k_2)} + \MxTwo{0 & B_y^\dag}{B_y^\dag & 0} e^{i(k_2-\khf)},
\end{multline}
\end{widetext}
where $M$ is a term odd under time-reversal (such as $\sigma^z$ or
$\tau^y$) and represents the added antiferromagnetic coupling in this
example.  The time-reversal operator takes the form
\begin{align}
	\TR & = -i\MxTwo{1_\tau \otimes \sigma^y & 0}{0 & 1_\tau \otimes \sigma^y} \mathcal{K}.
\end{align}

In the absence of $M$ this system also has two parity (spatial
inversion) centers, given by the operators:
\begin{align}
	\Pi_1 & = e^{i\frac{k_3}{2}} \MxTwo{0 & \tau^z}{\tau^z & 0}
	& \Pi_2 & = \MxTwo{\tau^z & 0}{0 & \tau^z}
\end{align}
The inversion center for $\Pi_1$ is between the two cubic sublattices
$X$ and $Y$, such that it swaps $X$ and $Y$. The inversion center for $\Pi_2$
is at $X$, such that it takes $Y$ to the next unit cell.
Their product results in a translation by half a unit cell: 
$\Pi_1 \Pi_2 = \Ta$.

\subsection{Construction from magnetically induced spin-orbit coupling}
\label{ModelAFSO}

\subsubsection{Motivation}

\begin{SCfigure}[20][tbp]
\begin{minipage}{45mm}
	\begin{align*}
	\xymatrix @!0 @R=10mm @C=16mm{
		&	\bullet \ar[dr]^{\vr_2}
			\save[] +<0mm,3mm> *{M_1} \restore
		\\	\bullet \ar[ur]^{\vr_1} \ar[dr]_{\vr_2}
			\save[] +<-3mm,0mm> *{X} \restore
		&&	\bullet
			\save[] +<3mm,0mm> *{Y} \restore
		\\&	\bullet \ar[ur]_{\vr_1}
			\save[] +<0mm,-3mm> *{M_2} \restore
	}
	\end{align*}
\end{minipage}
	\caption{%
		Four atoms placed in a rhombus configuration on the $xy$-plane
		The coupling between $X$ and $Y$ depends on the magnetization of $M_1$ and $M_2$.
	}
	\label{fig:RhombusAtoms}
\end{SCfigure}
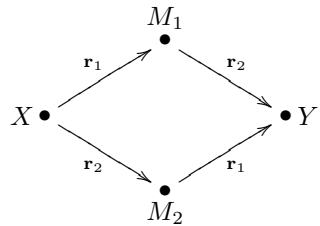

Consider four atoms placed in a rhombus geometry on the $xy$-plane as shown in Fig.~\ref{fig:RhombusAtoms}, with $X$ and $Y$ on opposite corners of the rhombus.
In the simplest model, the spin-orbit coupling term from $X$ to $Y$ 
is given by
$i\lambda_\textrm{SO} \sum \mathbf{d}_1 \times \mathbf{d}_2 \cdot \bm\sigma$,
where the sum is over the two paths
$X\rightarrow M_1 \rightarrow Y$, $X\rightarrow M_2 \rightarrow Y$, and
$\mathbf{d}_1, \mathbf{d}_2$ are the vectors along the bonds $X \rightarrow M_\ast$ and $M_\ast \rightarrow Y$ that the electron travels 
through.~\cite{HaldaneQHE88,KaneMeleQSH}
In this geometry this coupling vanishes as the cross products $\mathbf{d}_1 \times \mathbf{d}_2$ from the two paths cancel.

Now let $M_1$ and $M_2$ be magnetized in the $+z$ direction. This creates
a net magnetic field inside the rhombus breaking the symmetry between the
two paths $X \rightarrow M_\ast \rightarrow Y$. We can estimate its
orbital effect by attaching an Aharonov-Bohm phase $e^{\pm i \phi}$ to
each of the two paths, to produce a flux $2\phi$.
The coupling from $X$ to $Y$ now takes the form
\begin{multline}
	H_\textrm{SO} = i\lambda_\textrm{SO} \left[ e^{i\phi} \vr_1 \times \vr_2
			+ e^{-i\phi} \vr_2 \times \vr_1 \right] \cdot (c_Y^\dag \bm\sigma c_X)
		\\	\approx 2\phi\lambda_\textrm{SO} |\vr_2 \times \vr_1| (c_Y^\dag \sigma^z c_X),
\end{multline}
where we expect $\phi$ to be proportional to the $z$\hyp{}magnetization
of $M_\ast$.  Hence the magnetization of intermediate sites $M_1$ and
$M_2$ induces a spin-orbit interaction between the sites $X$ and $Y$.

If spins on $M_1$ and $M_2$ are aligned oppositely in the $\pm z$
direction, by contrast, there is no net magnetization in the rhombus and
the symmetry between the two paths $X \rightarrow M_\ast \rightarrow Y$
is restored. Rotating the system by $\pi$ along the axis through points
$X$ and $Y$, taking $M_1$ to $M_2$ and vice versa, we see that there are
no $\sigma^z$ couplings between the two sites.
Both cases are important in motivating the model to follow.

\subsubsection{Effective Hamiltonian}

We start with a rock-salt (FCC) structure with the conventional cubic unit cell of side length $1$.
In this setup there are four `$A$' sites located at $(0,0,0)$ and permutations of
$(\tfrac{1}{2},\tfrac{1}{2},0)$, while the `$B$' sites are located at
$(\tfrac{1}{2},\tfrac{1}{2},\tfrac{1}{2})$ and permutations of
$(0,0,\tfrac{1}{2})$.
The $B$ sites develop antiferromagnetic order along the $(111)$ planes
and magnetization in $\pm(1,1,1)$ direction.
In the antiferromagnetic state, the unit cell consists of four
layers: $A1$, $B\!\uparrow$, $A2$, and $B\!\downarrow$.

\begin{SCfigure}[20][htbp]
\begin{minipage}{40mm}
	\newcommand{\lab}[1]{ \save[] +<-3.5mm,0mm> *{#1} \restore }
	\begin{align*}
	\xymatrix @!0 @R=7mm @C=7mm{
		\bullet \lab{A1} &	&	\bullet \lab{A1}	&	&	\bullet \lab{A1}
	\\	&	\uparrow &	&	\uparrow \lab{B}
	\\	\bullet \lab{A2} &	&	\bullet \lab{A2}	&	&	\bullet \lab{A2}
	\\	&	\downarrow	&	&	\downarrow \lab{B}
	\\	\bullet &	&	\bullet	&	&	\bullet \lab{A1}
	}
	\end{align*}
\end{minipage}
	\caption{%
		Cross section of the model at (100) plane. The layers in a unit cell are $A1$, $B\!\uparrow$, $A2$, $B\!\downarrow$.
		Note that the magnetizations are not in-plane, but are only illustrated as such in this figure.
	}
	\label{fig:ABLayers}
\end{SCfigure}
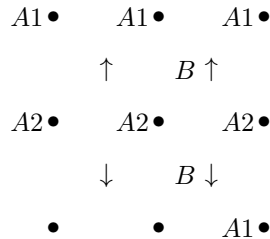

In this model there are spin up and spin down degrees of freedom at $A1$ and $A2$, but the electronic degrees of freedom at $B$ are eliminated, giving four ``orbitals'' per primitive cell.
The electrons hop between $A$ atoms by traveling through the magnetized $B$ sites, and we can see that there are always two such paths $A \rightarrow B \rightarrow A$ between adjacent $A$'s.
From Fig.~\ref{fig:ABLayers}, it is apparent that spin-orbit coupling between two $A1$'s on the same layer vanishes by our argument earlier, as the intermediate sites have opposite magnetization. In contrast, the spin-orbit coupling between $A1$ and $A2$ does not vanish.

Now we describe our model with the following hopping terms:
(1) spin\hyp{}independent hoppings between $A1$ and $A2$ atoms with
coefficient $t$,
(2) spin\hyp{}independent hoppings between $A$ atoms on the same plane with
coefficient $t'$,
and (3) spin-orbit term between $A1$ and $A2$ with effective coupling $\pm\lambda$.
As mentioned earlier, we take the energy to reside on $B$ sites as far above the energy scales $\lambda, t, t'$, effectively eliminating those degrees of freedom in our model.

We choose the primitive lattice vectors
\begin{align}
	\begin{bmatrix} \va_1 \\ \va_2 \\ \va_3 \end{bmatrix}
		= \begin{bmatrix} -\frac{1}{2} & 0 & \frac{1}{2} \\ 0 & -\frac{1}{2} & \frac{1}{2} \\ 1 & 1 & 0 \end{bmatrix}
			\begin{bmatrix} \va_x \\ \va_y \\ \va_z \end{bmatrix}
\end{align}
\begin{widetext}
in terms of the simple cubic basis $\va_x,\va_y,\va_z$.
The atoms $A1$ and $A2$ are placed at $-\frac{\va_3}{4}$ and $\frac{\va_3}{4}$ respectively within the unit cell.
Written in the basis $\ket{A_1^\uparrow}, \ket{A_1^\downarrow},
\ket{A_2^\uparrow}, \ket{A_2^\downarrow}$, the Hamiltonian takes the
form
\newcommand{\tkhf}{ \tfrac{k_3}{2} }
\begin{align}
	H &= \MxTwo{T' & U^\dag}{U & T'},
	&&\textrm{where}\quad \begin{aligned}
		T' & = 2t' \big[ \cos(k_1) + \cos(k_2) + \cos(k_1 - k_2) \big] \mathds{1} ,
	\\	U & = 2t \big[ \cos(\tkhf) + \cos(k_1+\tkhf) + \cos(k_2+\tkhf) \big]
			\\	&\quad\quad - 2i\lambda \big[ \sin(\tkhf)\sigma^z + \sin(k_1+\tkhf) \sigma^x + \sin(k_2+\tkhf) \sigma^y \big] ,
	\end{aligned}
	\label{HamiltonianAFSO}
\end{align}
\end{widetext}
which is gapped (in the bulk) when $|t'| < |t|,\frac{1}{\sqrt{3}}|\lambda|$.
The time-reversal operator has the representation
\begin{align}
	\TR & = -i\MxTwo{\sigma^y & 0}{0 & \sigma^y} \mathcal{K}.
\end{align}
We are interested in the regime where $t'$ is much smaller than $t$
and $\lambda$, as this leads to a gap in the surface spectrum also.
Unfortunately, we cannot provide a good argument why $t'$ (in-plane
hopping) should be much less than $t$ (interplane hopping) in a real
material.

This model has spatial inversion symmetry, given by the operator
$\Pi = \SmallMxTwo{0 & \mathds{1}}{\mathds{1} & 0}$, which in effect
swaps the layers $A1$ and $A2$.  The filled bands at the momenta
$(k_1,k_2,k_3) = (0,0,0), (0,\pi,0), (\pi,0,0), (\pi,\pi,0)$ have parity
$-1,-1,-1,+1$ respectively, so the model is in the non-trivial
topological phase.

In this model, $\lambda$ is related to the parameter breaking
time-reversal symmetry, at the same time protecting the bulk gap.
If we turn the
parameter $\lambda$ down to zero, we will not get a STI at
$\lambda=0$, rather, the model becomes conducting.

\section{Surface band structure}

The bulk electronic band structure of an AFTI must be gapped to allow the
topological distinction between the trivial phase and the non-trivial
phase.  At the boundary between domains of two topologically distinct
phases we typically expect a gapless surface spectrum, as is the case at the
edges of quantum Hall and quantum spin Hall systems, as well as at the 
surfaces of the STI (vacuum is in the trivial phase).
However, it should be noted that this is not strictly necessary.  For 
example, while time-reversal symmetry requires doubly degenerate 
states, leading to gapless boundary modes between topological phases, it
is known that breaking time-reversal but preserving inversion can give a
topological phase whose surface states are gapped.~\cite{TurnerTopoEntanglement}  

We distinguish between two classes of surfaces, depending on the
plane of the cut relative to the crystal structure. We call a surface
\textbf{type~F}(erromagnetic) if it breaks the $S$ symmetry in the bulk, and
\textbf{type~A}(ntiferromagnetic) if it preserves the symmetry.
Heuristically, the distinction can be visualized by imagining a
ferromagnetic/antiferromagnetic moment with a Zeeman coupling to the electron's
spin, as in model~A above.
Then a type~F surface will have all its spins aligned, and a net
magnetization on the surface. A type~A surface will have
antiferromagnetic order such that we can always choose the primitive lattice vector $\va_3$
parallel to the surface.
As an example: In model~A with staggered magnetization on a cubic lattice, $\{111\}$ planes are type~F, while planes $\{110\}$ and
$\{100\}$ are type~A.

There are an odd number of Dirac cones on a clean type~A surface,
analogous to the STI.
We can see why the surface (parallel to $\va_3$) is
gapless by looking at the $k_3=0$ line on the surface spectrum,
which is the boundary of the $k_3=0$ plane in the bulk BZ.
Since the plane carries a non-trivial topological (QSH) phase, its
boundary must be gapless.

The Dirac cone's stability may also be explained by looking at a
constant energy curve $\gamma$ in the surface spectrum. This curve
must be its own time-reversal image because of the symmetry between
$\vk$ and $-\vk$.
The Berry phase of this curve $\phi = \oint_\gamma\! \Tr{A}$ is ambiguous by
integer multiples of $2\pi$, so $S$ symmetry forces this to be $0$
or $\pi$, for the same reason it forced $\theta = \pi$ in Sec.~\ref{MagnetoelectricCS}.
As in the STI, a $\pi$ phase implies that the Fermi surface encloses
an odd number of Dirac cones.
However, any defect or impurity will break the translational and $S$ symmetry
on the surface, thereby opening a gap.  This is analogous to the
effect of a magnetic defect on the surface of a STI.

For a type~F surface, $S$ symmetry is broken on the surface and the
usual protection for Dirac cones or conducting surfaces no longer
exists.  If the bulk and surface spectrum are fully gapped
(\textit{i.e.}, not a semi-metal), then the surface will exhibit the
half-integer quantum Hall effect, to be discussed in the next
section.

\begin{figure}[tbp]
	\centering
	\includegraphics[width=0.9\columnwidth]{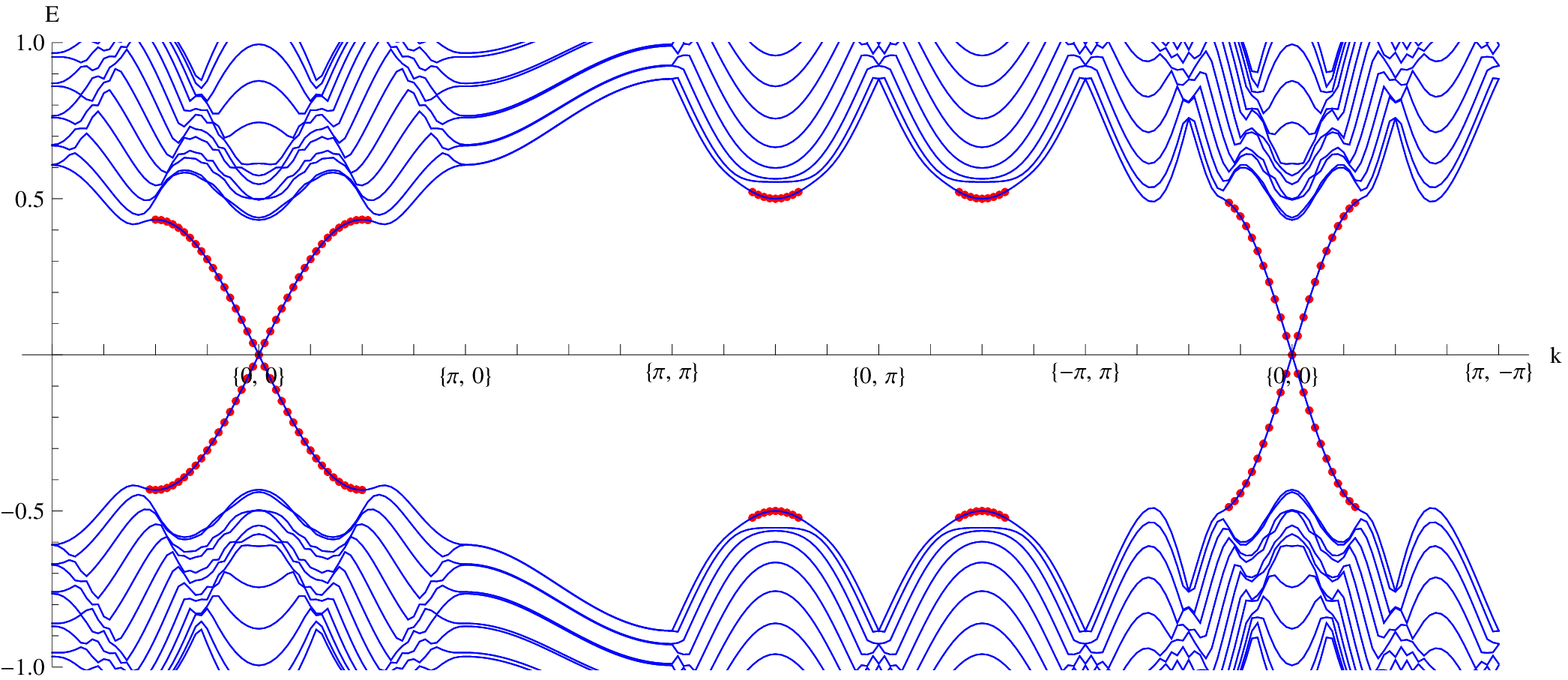}
	\\ \vspace{1ex}
	\includegraphics[width=0.9\columnwidth]{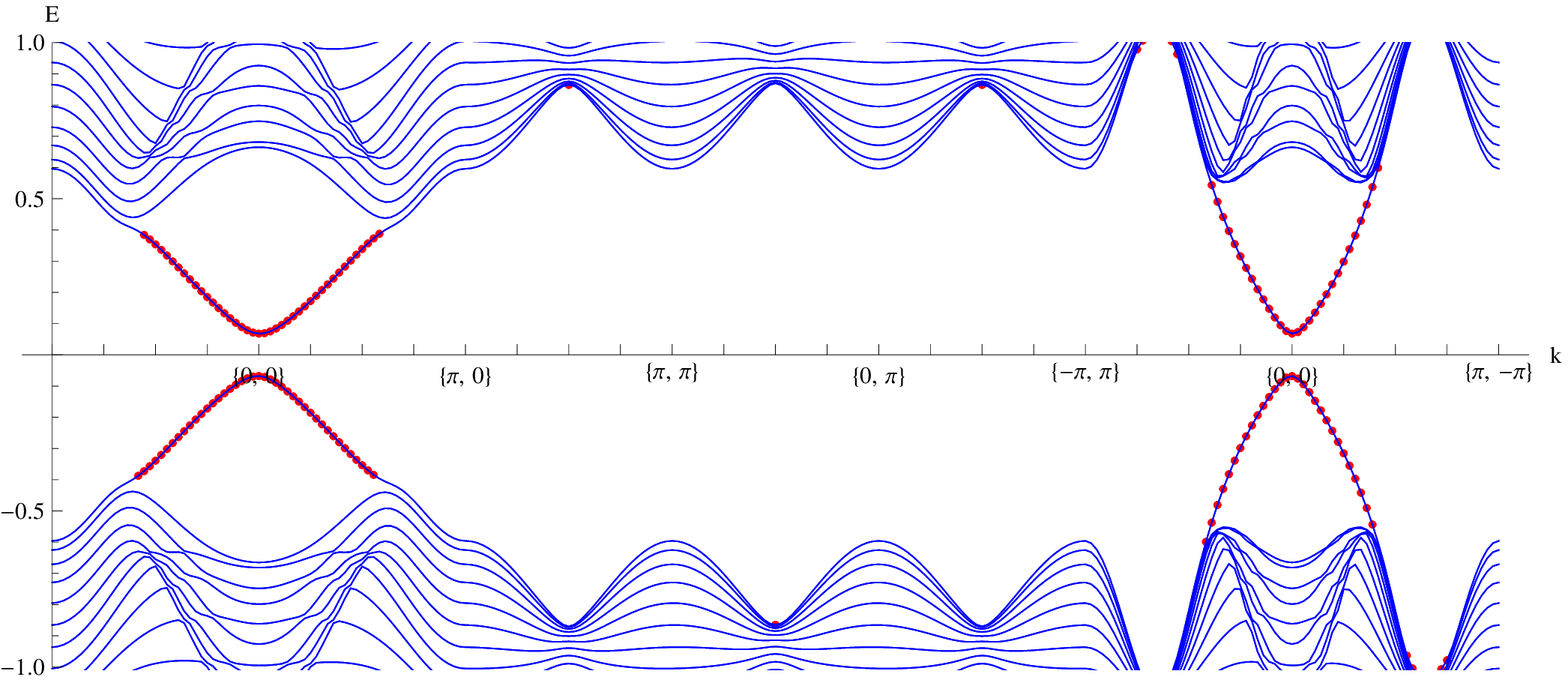}
	\caption{%
		(Color online)
		The bulk and surface band structure for model~A, along the $(100)$ (type~A, parallel to $\va_2,\va_3$) and $(\bar{1}\bar{1}1)$ plane (type~F, parallel to $\va_1,\va_2$) respectively.
		The red dots indicate surface modes.
		The parameters used are: $v = 0.5, m = 2, t = 1, M = \sigma^z$ with 13 layers.
	}
	\label{fig:ModelABand}
\end{figure}

In Fig.~\ref{fig:ModelABand}, we present the band structure of
model~A for slabs with type~A and type~F surfaces. Since this model
is built from a STI, the band structures are similar.~\cite{FuKaneMeleTI3D}

\begin{figure}[tbp]
	\centering
	\includegraphics[width=0.9\columnwidth]{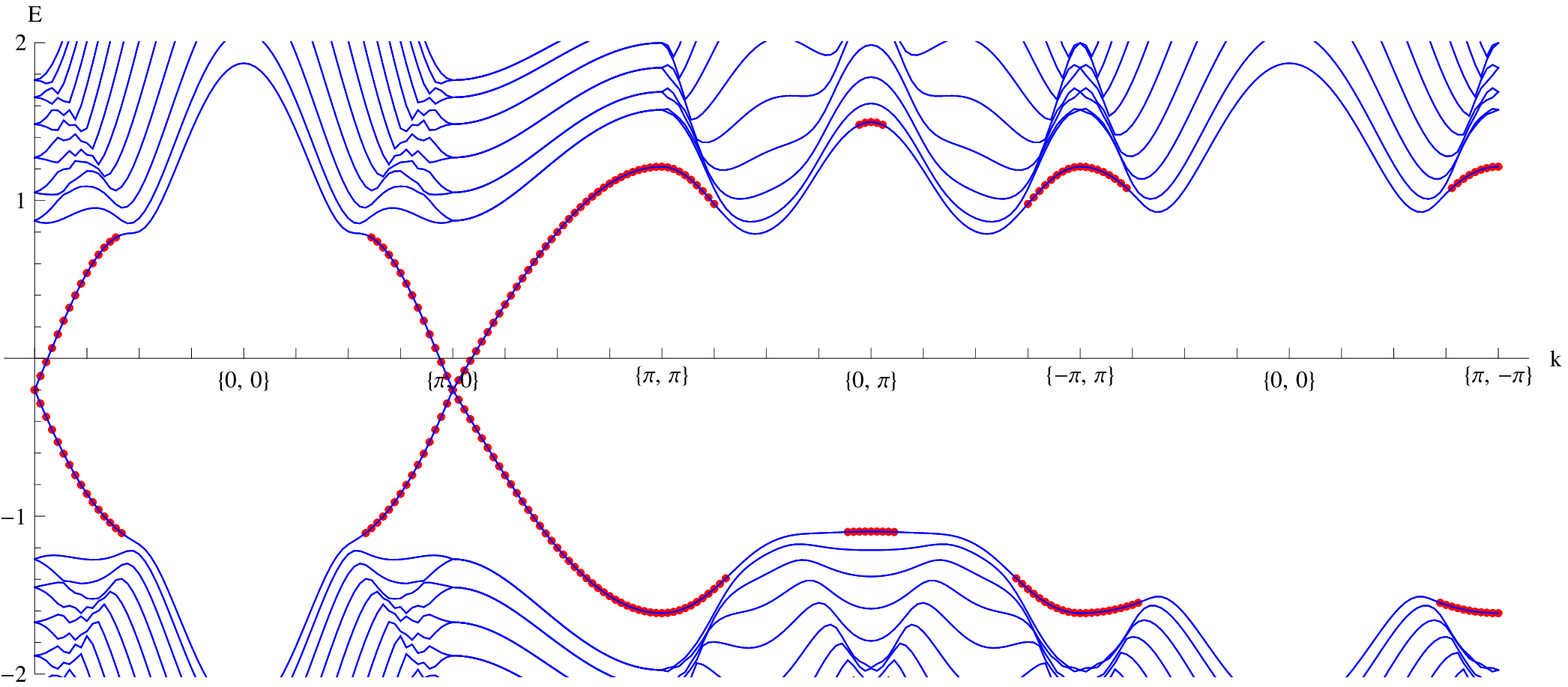}
	\\ \vspace{1ex}
	\includegraphics[width=0.9\columnwidth]{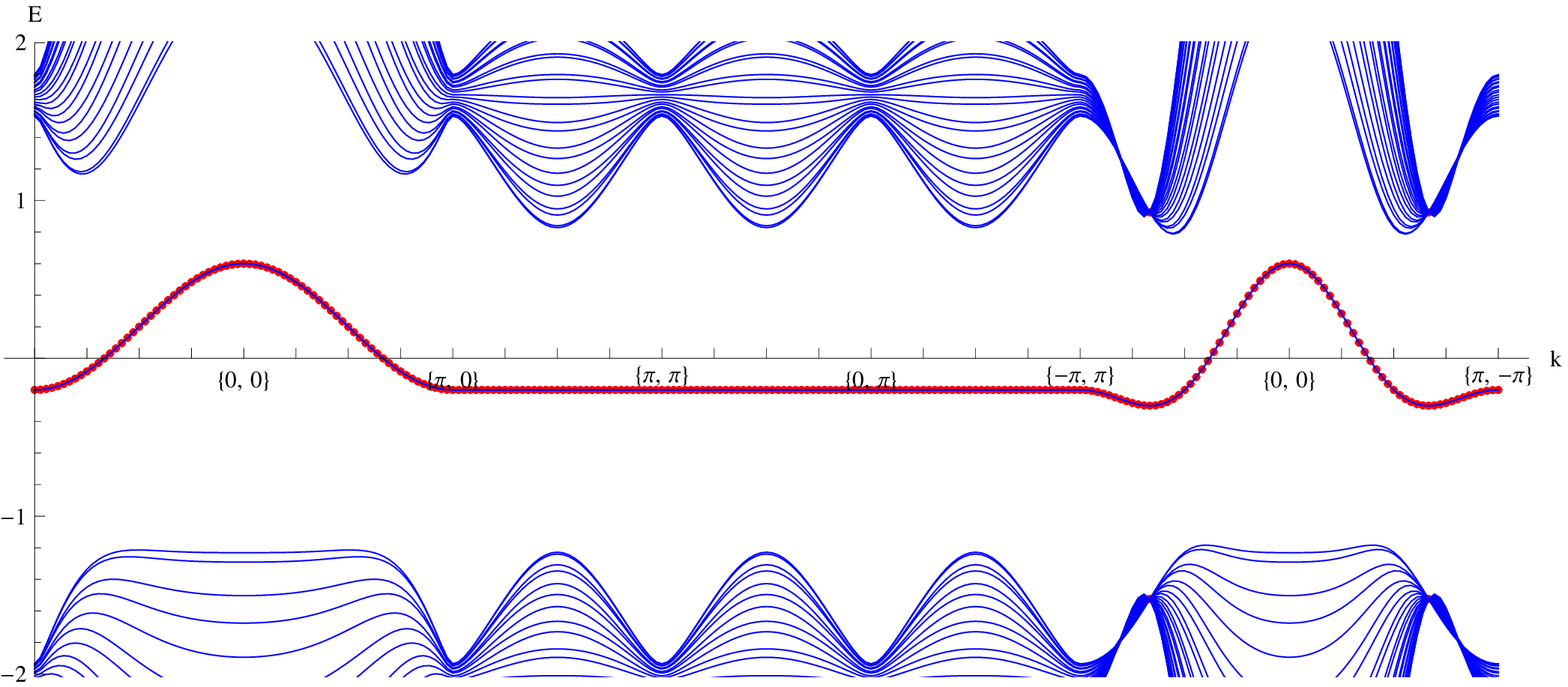}
	\caption{%
	(Color online)
		The bulk and surface band structure for model B, along the $(1\bar{1}\bar{1})$ plane (type~A, parallel to $\va_2,\va_3$) and $(111)$ plane (type~F, parallel to $\va_1,\va_2$) respectively.
		The red dots indicate surface modes.
		The parameters used are: $\lambda = 0.5, t = 1, t' = 0.1$ with 13 layers.
	}
	\label{fig:ModelBBand}
\end{figure}

For model~B, the surface parallel to $\va_1$ and $\va_2$ is type~F and
its excitations are exactly solvable, with dispersion
\begin{align}
	E_s(k_1,k_2) = 2t' \big( \cos k_1 + \cos k_2 + \cos (k_1-k_2) \big).
\end{align}
As the dispersion shows, a surface spectrum exists for all values of
$t'$ and non-zero values of $t$ and $\lambda$.  This model has the
peculiar feature that the surface spectrum is completely disconnected
from the bulk, that is, it forms a complete two\hyp{}dimensional band
structure.  Figure~\ref{fig:ModelBBand} shows the bulk and surface
band structure for two different cuts.  In the $(111)$ cut, a small $t'$
is desired if we want to avoid band overlaps between the valence,
conductance, and surface spectrum, giving us an insulator.

In the presence of a sufficient number of random defects, we expect that the surface electronic states are described by the unitary symmetry class because of the broken time-reversal symmetry.
  That symmetry class only has extended states at isolated values of the chemical potential; in general the surface state will have zero diagonal conductivity, with half-integer quantum Hall plateaus.  The transitions between these plateaus appear when the chemical potential passes through an extended state.  These transitions can be regarded as a realization of the two\hyp{}dimensional quantum Hall effect in zero net field discussed by Haldane.~\cite{HaldaneQHE88}  Note that since both top and bottom surfaces of a slab will have half-integer plateaus, the total quantum Hall effect when diagonal conductivity is zero is always integral, as required for a single\hyp{}electron two\hyp{}dimensional system.

\section{Ferromagnetic surfaces and half-integer quantum Hall effect}

In this section we present two perspectives on the half quantum Hall effect on type~F surfaces, along with numerical calculations to justify our claim.

If one views the antiferromagnet as a STI with time-reversal breaking term opening a surface gap,
then the half QHE can be viewed as the root of the bulk magnetoelectric coupling $\theta = \pi$.
This effect follows from the gapped Dirac dispersion of the surface states.
The sign of the Hall conductance depends on
the sign of the effective Dirac mass,~\cite{Jackiw84,Ishikawa84,HaldaneQHE88}
which here is set by the
direction of the Zeeman field at the surface.

An alternate perspective of the AFTI surface comes from a comparison 
to the quantum spin Hall effect.
As described in Sec.~\ref{SectionZ2Inv}, the $\ZZ_2$ invariant is computed from the two\hyp{}dimensional plane $k_3=0$ and the symmetry operator 
$S|_{k_3=0}$ in precisely the way that the quantum spin Hall (QSH) 
invariant is computed from the two\hyp{}dimensional BZ and $\TR$.
The QSH insulator may be constructed by combining two copies of a QH layer with opposite spin and Chern number $\pm n$.  Time-reversal takes one layer to the other, making the combination of the two $\TR$-invariant.
In reality spin is rarely conserved, allowing the two layers to mix, making the Chern number of each spin ill-defined.
However, a residual $\ZZ_2$ topological invariant remains,~\cite{KaneMeleQSH,KaneMeleZ2} and we can consider the QSH as being adiabatically connected to the two-QH-layer model, but with the topological invariant $n$ only preserved $\bmod 2$.

\begin{SCfigure}[1.25][htb]
	\includegraphics[width=0.4\columnwidth]{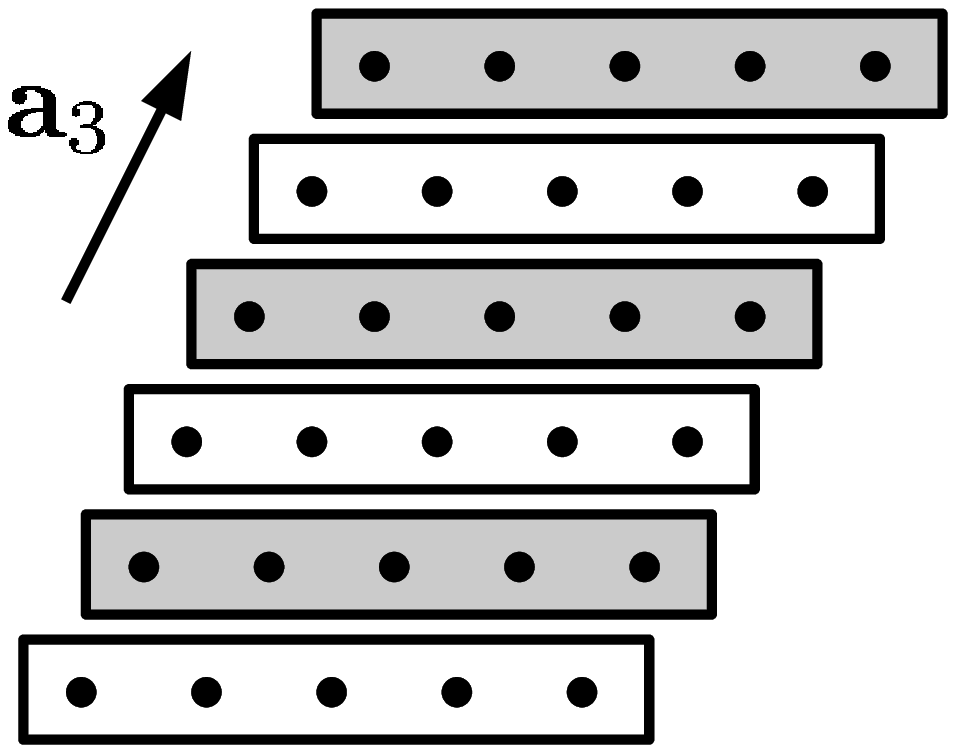} \vspace{5mm}
	\hspace{2ex}
	\caption{%
		Construction of antiferromagnetic topological insulator by staggering quantum Hall layers.
		The shaded and unshaded boxes represents Chern number of $\pm1$.
		The left and right surfaces are type~A and gapless, while the top and bottom surfaces are type~F with half\hyp{}quantum Hall effect.
	}
	\label{fig:LayeredQHE}
\end{SCfigure}
By analogy, we can construct an AFTI by stacking QH layers, with alternating Chern number of $\pm1$ (Hall conductivity $\pm e^2/h$), as shown in Fig.~\ref{fig:LayeredQHE}.
The ``up'' ($+1$) layers are related to the ``down'' ($-1$) layers by $S$ symmetry, hence they are spatially offset from one another.
Just like the QSH case, we can expect the layers to couple to one another, in a way that makes the Chern number ill-defined on a per-layer basis. Once again, it is appropriate to consider the AFTI to be adiabatically connected the staggered QH layer model.
In the stacked QH model, the Hall conductance in the bulk averages to zero, as the conductance of any individual layer is cancelled by neighboring layers of opposite type. In other words, any long-wavelength probe of the system will be unable to discern the individual QH layers. However, the QH layers at either end of the stack are not completely cancelled, there is a half QHE at both surface.

To confirm this
picture, we can consider a slab with type~F surfaces and compute the
2D Hall conductivity as a function of position (layer).  In units
of $e^2/h$, the (two\hyp{}dimensional) conductivity in layer $n$ can be
computed from~\cite{EssinMPAxion}
\begin{align}
	C(n) = \frac{i}{2\pi} \int \operatorname{tr} \left[
			\mathcal{P} (d \mathcal{P}) \wedge \tilde{\mathcal{P}}_n (d \mathcal{P})
		\right].
	\label{layerConductance}
\end{align}
Here $\mathcal{P} = \sum_{\textrm{occ}} \ket{u_\vk} \bra{u_\vk}$ is the projector onto
occupied wave functions at $\vk$ and $\tilde{\mathcal{P}}_n$
is the projector onto basis states localized in layer $n$.

\begin{figure}[hbt]
	\includegraphics[width=0.8\columnwidth]{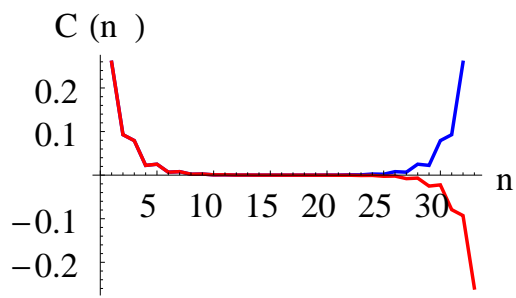}
	\caption{%
		(Color online)
		Hall conductance spatially resolved for 32 (blue) and 33 (red) layers of model~B.
		The Fermi level is set below zero to include only the bulk valence modes and no surface modes.
		The parameters used for this plot are: $\lambda = 0.5, t = 1, t' = 0.1$.
	}
	\label{fig:HallConductance}
\end{figure}
Figure~\ref{fig:HallConductance} shows the results of such a computation
on a slab cut from the rock-salt model (B) introduced in
Eq.~\eqref{HamiltonianAFSO} with type~F surfaces.
In this model, when the Zeeman field on opposite surfaces points in
opposite directions (blue, upper curve) the total conductance of the slab is
$C = 1$, with each surface having a net $C = 1/2$; adding a
layer such that the two surfaces have the same Zeeman field switches
the conductance on that surface from $+1/2$ to $-1/2$, so that the
total slab conductance vanishes.  Note that
the total conductance of a slab is always an integer, as
required.~\cite{TKNN,ASSHomotopy,HaldaneQHE88}

\begin{figure}[bth]
	\includegraphics[width=0.8\columnwidth]{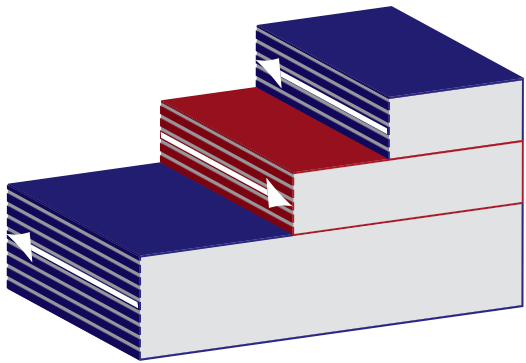}
	\caption{%
		(Color online)
		1D quantum wire on type~F surface step edge.
		The red and blue regions represent ferromagnetic layers magnetized in opposite directions.
		There is a gapless chiral quantum wire at each step edge, with chirality indicated by the arrow at the edge.
	}
	\label{fig:StepEdge}
\end{figure}

Now, at the interface between two integer quantum Hall domains whose
conductance $C$ differs by 1, there will be a chiral boundary mode
with conductance $e^2/h$, which can be thought of as ``half a quantum
wire.''  In the situation outlined above, putting the two slabs with
different conductance together is equivalent to making a slab with a
step edge on one surface, and the chiral mode will reside at this step
edge.  Such a mode should give an observable signature in a tunneling
experiment (Fig.\ref{fig:StepEdge}).

It is natural to ask, what if one rotates the antiferromagnetic moment by
$\pi$, flipping all the spins and effectively ``peeling'' off a layer of
type~F surface?
Since the sign of the surface conductance $C$ changes during this process, the surface (or bulk) gap must close at some magnetization orientation.
This is analogous to applying a magnetic field to a STI surface. For $\mathbf{B}$ parallel to the surface, the Dirac cone shifts in momentum space but no gap opens. Any infinitesimal component of $\mathbf{B}$ out of the plane will open a gap, hence going from $\mathbf{B}$ out of the surface to $\mathbf{B}$ into the surface must necessarily close the surface gap. (In model~B, the bulk gap would close while rotating the magnetization.)

\section{Conclusions and possible relevance to $\textrm{GdBiPt}$}

In this paper, we have looked at the topological classification of 
materials breaking both time-reversal $\TR$ and translational 
symmetry $\Ta$, but preserving the combination $S = \TR\Ta$, and 
found a $\ZZ_2$ classification within the $S$ symmetry class that 
leads to the existence of an antiferromagnetic topological insulator
(AFTI).  In the most basic picture, an AFTI can be obtained from 
adding a staggered magnetization to a strong topological insulator 
(STI).  Macroscopically, $S$ symmetry implies a quantized 
magnetoelectric response 
$\frac{\partial P}{\partial B}\big\vert_{B=0} = \frac{\theta}{2\pi}\frac{e^2}{h}$
with $\theta = \pi$ for an AFTI.
We have also demonstrated that the surface spectrum depends on the 
surface cut, classified as type~A/F.  Type~A surfaces possess an 
antiferromagnetic order that preserves $S$ symmetry, with associated
gapless excitations that can be gapped by disorder.  Type~F surfaces 
break $S$ symmetry and are typically gapped, analogously to the 
situation of a Zeeman field on the surface of a STI.
The new AFTI state is topological in a weaker sense than the 
strong 3D topological insulator, because its surface state is 
dependent on the surface plane and not generally stable to disorder; 
in that respect it is similar to the weak topological insulator 
in 3D or the ``Hopf insulator.''~\cite{MooreHopfIns}  (The number of 
Dirac cones in a STI also depends on the surface plane, but there is 
always an odd number of such cones.)

The magnetoelectric coupling $\theta=\pi$ requires the half quantum 
Hall effect at the surface, provided the surface spectrum is gapped. 
Our numerical calculations based on explicit band models agree with 
these results.  Finally, we predict the existence of chiral 1D 
quantum wires at type~F surface step boundaries, an experimental 
signature verifiable via scanning tunneling measurements. 

The recent proposals that many Heusler compounds may be topological insulators,~\cite{HasanHeusler,FelserHeusler}
together with the antiferromagnetic order in GdBiPt below 9 K,~\cite{CanfieldAFGdBiPt}
suggest a possible candidate for the state proposed here.
Transport experiments indicate that GdBiPt is a semiconductor with a narrow gap.~\cite{CanfieldAFGdBiPt,MatsubaraPtGdBi}
The Gd sites form an FCC lattice and hence their antiferromagnetic interaction is frustrated, and further experiments (\textit{e.g.} neutron scattering) are required to determine if the antiferromagnetic order falls under the $S$ symmetry class described in this paper.
At least one related Heusler antiferromagnet (MnSbCu) is known to have antiferromagnetic ordering [alternating spin directions on $(111)$ planes] which belongs in the $S$ symmetry class.~\cite{ForsterJohnstonCuMnSb68}
Should the material be truly insulating (\textit{i.e.}, have a bulk gap) through its antiferromagnetic transition, it suffices in principle to check if it is a strong topological insulator above the N\'eel temperature.

We have provided a topological classification and experimental consequences for a particular combination of time-reversal symmetry and a lattice symmetry ($\TR \Ta$).  Other such combinations of time-reversal and crystal symmetries could lead to new topological materials beyond those in the exhaustive classification of topological insulators stable to disorder.~\cite{SFRLClassification,KitaevClassification}

\section{Acknowledgements}
The authors gratefully acknowledge discussions with S.~Ryu, O.~Yazyev, and D.~Vanderbilt.
The work was supported by NSF under Grant No. DMR-0804413 (R.M. and J.E.M.) and WIN (A.E.).

\appendix

\section{Invariance to choice of unit cell}
\label{SecUnitCellChange}

The construction of the $\ZZ_2$ invariant in Sec.~\ref{SectionZ2Inv} required a certain choice
of unit cell.
In this section, we will demonstrate that different
choices of the unit cell will yield the same result.  In particular,
we show that different ways to choose the doubled unit cell are
equivalent given a choice of structural cell.

Begin with a Hamiltonian $H$ defined for a set of primitive
translation vectors $\va_i$, along with the operators $\Ta$ such that
$\Ta^2$ translate by $-\va_3$.  We can always divide the Hilbert
space in to two subspaces: $X$ and $Y$, such that the translation
operator $\Ta$ takes $Y$ to $X$, and $X$ to the $Y$ in another unit
cell.  Physically $X$ and $Y$ represent the structural unit cell whose
symmetry is broken by antiferromagnetism.

Construct a new unit cell by leaving $X$ fixed but taking $Y$ from a
cell $\vR$ relative to the original. In the new system the lattice
vector $\tilde{\va}_3 = \va_3 + 2\vR$ such that $\tildeTa^2$
translates the system by $-\tilde{\va}_3$.
We want to show that the $\ZZ_2$ invariant calculated for the new
Hamiltonian ($\tilde{H}$ on the $\tilde{k}_3=0$ plane) is identical
to that of the original one ($H$ on the $\tilde{k}_3=0$ plane).

\begin{figure}[tbp]
	\subfigure[]{
		\label{fig:CellChange-a}
		\includegraphics[width=0.35\columnwidth]{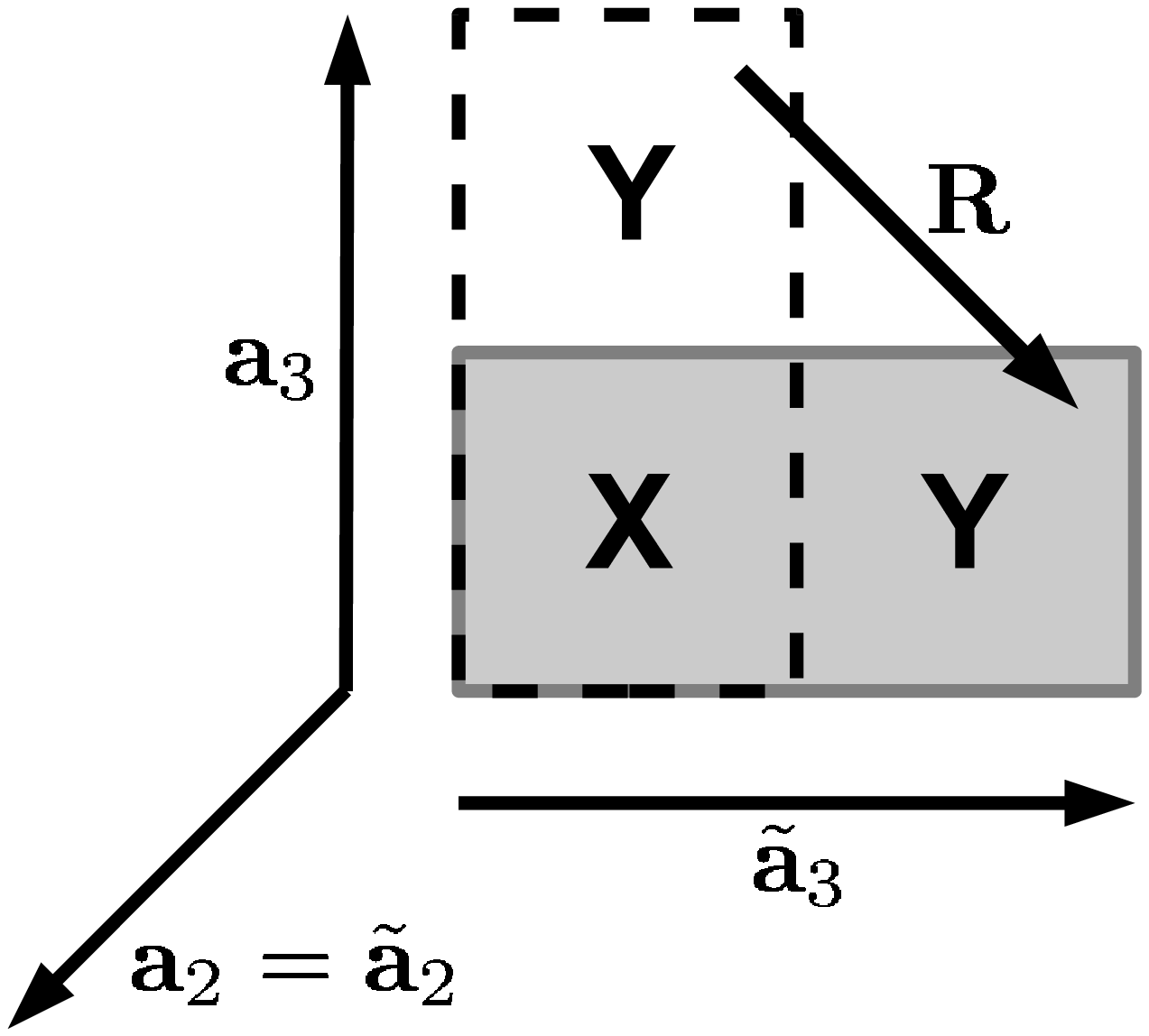} }
	\hspace{1ex}
	\subfigure[]{
		\label{fig:CellChange-b}
		\includegraphics[width=0.45\columnwidth]{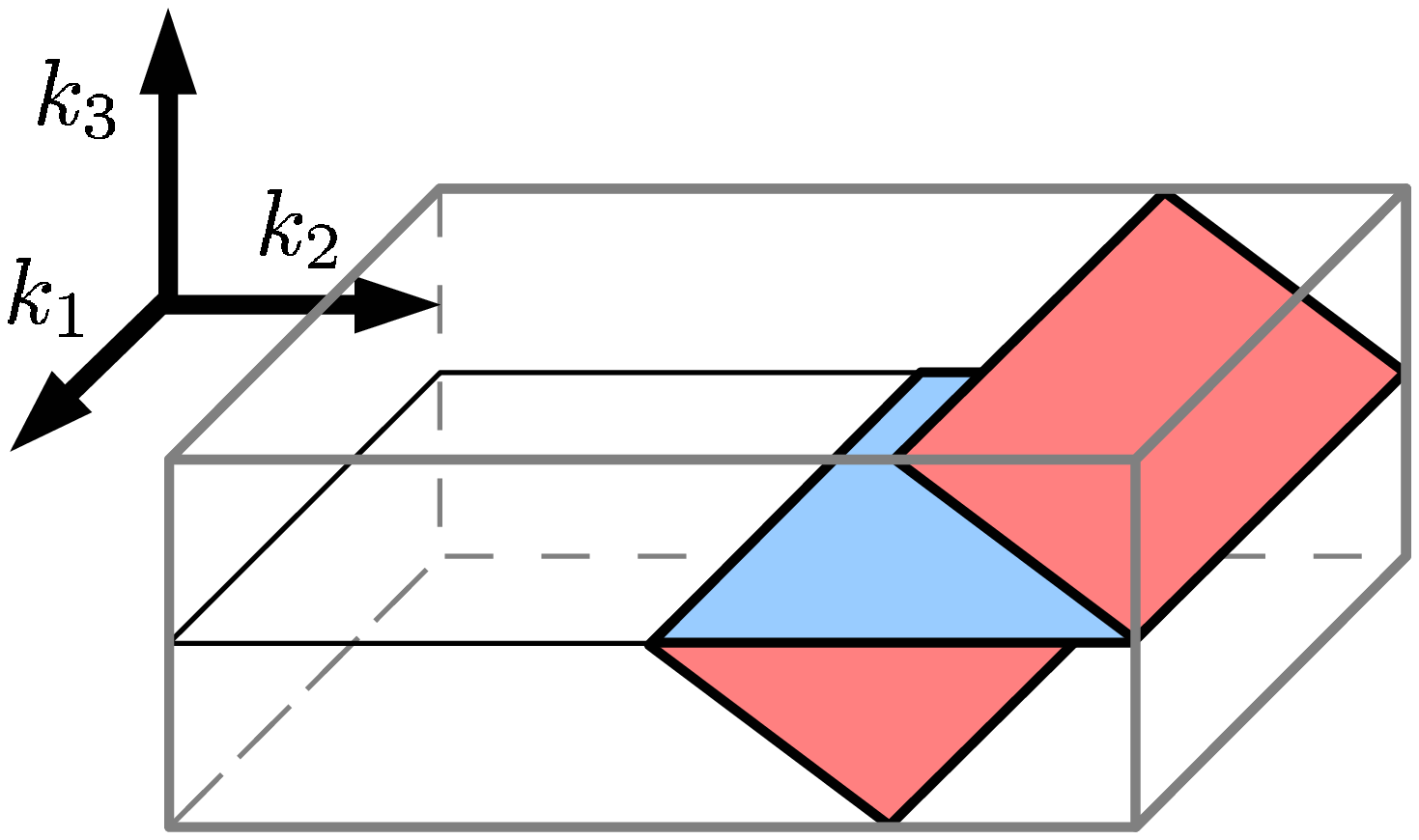} }
	\\
	\subfigure[]{
		\label{fig:CellChange-c}
		\includegraphics[width=0.9\columnwidth]{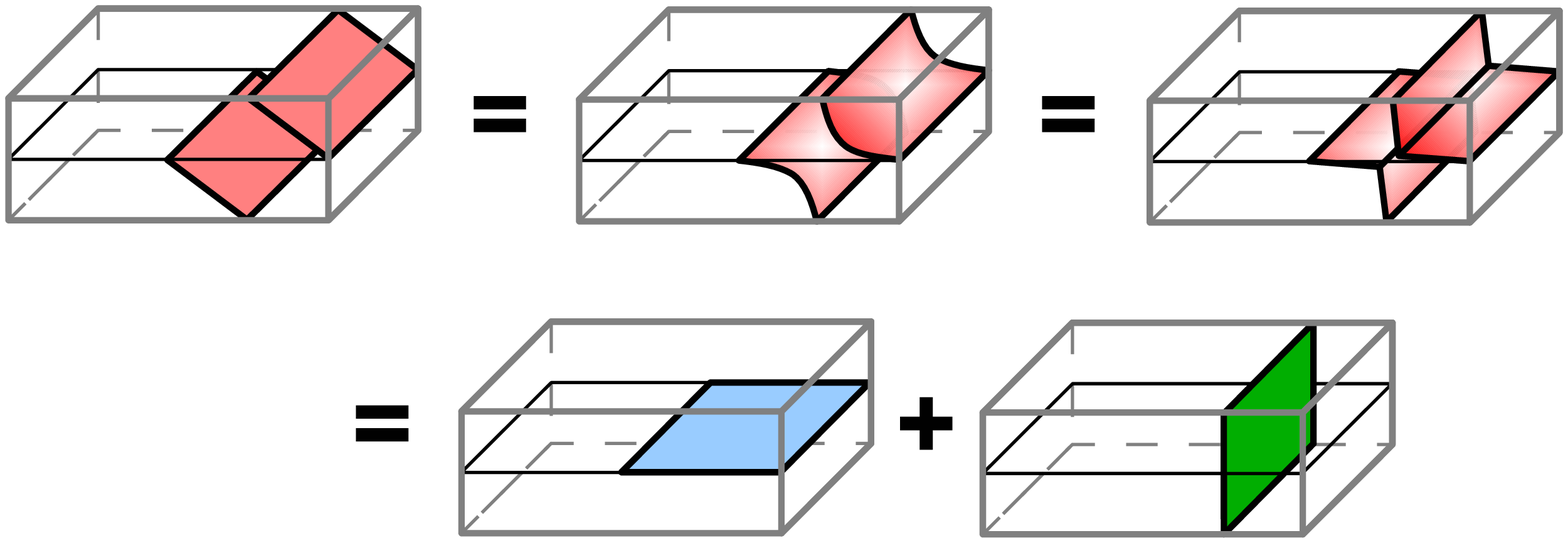} }
	\caption{%
		(Color online) Example of changing the unit cell.
		(a) The original cell (rectangle with dashed border) is transformed to the new cell (shaded rectangle with solid border) by keeping the $X$ portion fixed and changing $Y$, where $\vR = -\va_2-\va_3$ is the displacement vector.
		In this example, the original vectors $\va_2 = -\hat{x}-\hat{y}$, $\va_3 = 2\hat{y}$, and $\va_1$ points out of the plane. The new vectors $\tilde{\va}_3 = \va_3 + 2\vR = -2\va_2-\va_3 = 2\hat{x}$ and we choose $\tilde{\va}_1$ and $\tilde{\va}_2$ to remain fixed.
		(b) The Brillouin zone. The blue plane is the effective Brillouin zone (EBZ) for $k_3=0$, and the red plane is the EBZ for $\tilde{k}_3=0$.
			The $\ZZ_2$ invariant computed for these two planes are the same.
		(c) Deformation of the new EBZ (red), which decomposes into the old EBZ (blue) and a boundaryless plane (green).
	}
\end{figure}

Here we remind the reader of the method used in this section to compute
the $\ZZ_2$ topological invariant.~\cite{FuKaneTRPZ2,MooreBalents06}
First we pick an ``effective Brillouin zone'' (EBZ) which is half of the Brillouin zone such that
time-reversal will map it to the other half. The boundary of EBZ must
be time\hyp{}reversal image of itself.  The element of $\ZZ_2$ is computed
by the integrating the connection and curvature:
\begin{align}
	D = \frac{1}{2\pi} \left[\;
				\oint_{\partial\textrm{EBZ}}\hspace{-2ex} \mathcal{A}
				- \int_{\textrm{EBZ}}\hspace{-1ex} \mathcal{F}
		\;\right] \,\bmod 2 ,		\label{ChernParity}
\end{align}
where the [$U(1)$] connection $\mathcal{A} = \sum_\textrm{occ} \bra{u} id \ket{u}$ is summed over occupied bands
and curvature $\mathcal{F} = d\mathcal{A}$ is its exterior derivative in momentum space.
The curvature $\mathcal{F}$ is ``gauge invariant'' (does not depend
on the choice of basis functions for occupied states), but $\mathcal{A}$
depends on a particular choice of gauge for the wave functions. The
boundary integral in the formula above requires that the
wave functions at $\vk$ and $-\vk$ be $S$\hyp{}conjugate pairs.  Any
choice of the EBZ will give the same $\ZZ_2$ invariant.

The effect of the coordinate transformation $k_1,k_2,k_3 \rightarrow \tilde{k}_1,\tilde{k}_2,\tilde{k}_3$
changes the EBZ on which we compute the topological invariant.
Since the momentum variables are related by $\tilde{k}_3 = k_3 + 2\vR\cdot\vk$,
we can always choose the EBZ for the new and old systems such that
they share a common boundary, namely, the two lines satisfying
$\vR\cdot\vk \in \{0,\pi\}$.
This guarantees that the boundary
integral terms ($\oint\!\mathcal{A}$) in (Eq.~\ref{ChernParity}) are identical in the two cases.

As for the term integrating curvature over the EBZ, we can try to deform
the new EBZ to match the old EBZ.
This deformation is allowed by the fact that $\mathcal{F}=d\mathcal{A}$
is a closed 2-form; any local deformation to the surface (\textit{i.e.},
one that preserves $\mathcal{A}$ on the boundary) will preserve the
integral $\int\!\mathcal{F}$.
As Fig.~\ref{fig:CellChange-c} shows, we cannot always deform one EBZ
to the other; however,
we can always
decompose the new EBZ into the old EBZ plus planes with no boundaries.
These closed planes which are either contractible, or they span a torus
in the Brillouin zone.
$S$ symmetry requires that the Chern number vanishes on
all closed two\hyp{}dimensional surfaces, and it follows that the integral (Eq.~\ref{ChernParity})
evaluates to the same value for new and old unit cell. In other
words, the $\ZZ_2$ invariant does not depend on how we choose the
unit cell.

We can also view the $\ZZ_2$ invariant as an obstruction to finding a continuous basis (along with the appropriate Bloch periodic boundary  conditions) for the wave functions respecting $S$ symmetry in the
entire Brillouin zone.~\cite{FuKaneTRPZ2} The material is in a
trivial phase if such a basis exists.  This intepretation is much
harder to ``compute'' then the original definition, but is powerful in
what it implies.  For example,
any (single-valued) unitary transformation or a change of coordinates will not affect the obstruction of finding such basis, and
it is rather straightforward from the definition that the $\ZZ_2$
invariant is independent of unit cell choice.

\bibliography{AFTI}

\end{document}